\documentstyle[12pt,psfig]{article}

\newcommand{\gsim}{\lower.7ex\hbox{$\;\stackrel{\textstyle>}{\sim}\;$}}
\newcommand{\lsim}{\lower.7ex\hbox{$\;\stackrel{\textstyle<}{\sim}\;$}}

\newcommand{\met}{\rlap{\,/}E_T}
\newcommand\beq{\begin{equation}}
\newcommand\eeq{\end{equation}}
\newcommand\bear{\begin{eqnarray}}
\newcommand\eear{\end{eqnarray}}
\def\ap  #1 #2 #3 #4 {Ann.~Phys.         {\bf  #1}, #2 (#3)#4 }
\def\aplb#1 #2 #3 #4 {Acta Phys.~Pol.    {\bf B#1}, #2 (#3)#4 }
\def\cpc #1 #2 #3 #4 {Comp.~Phys.~Comm.  {\bf  #1}, #2 (#3)#4 }
\def\jetp#1 #2 #3 #4 {JETP Lett.         {\bf  #1}, #2 (#3)#4 }
\def\npb #1 #2 #3 #4 {Nucl.~Phys.        {\bf B#1}, #2 (#3)#4 }
\def\plb #1 #2 #3 #4 {Phys.~Lett.        {\bf B#1}, #2 (#3)#4 }
\def\prd #1 #2 #3 #4 {Phys.~Rev.         {\bf D#1}, #2 (#3)#4 }
\def\prl #1 #2 #3 #4 {Phys.~Rev.~Lett.   {\bf  #1}, #2 (#3)#4 }
\def\ptp #1 #2 #3 #4 {Prog.~Theor.~Phys. {\bf  #1}, #2 (#3)#4 }
\def\zpc #1 #2 #3 #4 {Zeit.~Phys.        {\bf C#1}, #2 (#3)#4 }

\begin{document}

\baselineskip=24pt

\thispagestyle{empty}
\begin{titlepage}
\title {\vspace{-2.0cm} 
\hfill {\normalsize FERMILAB--PUB--99/078--T}\\
\vspace{0.9cm} 
Supersymmetry Reach of the Tevatron via Trilepton,\\
Like-Sign Dilepton and Dilepton plus Tau Jet Signatures }

\vspace{3cm}

\author{\\
{\sc Konstantin T. Matchev}\\
{\small Theoretical Physics Department}\\
{\small Fermi National Accelerator Laboratory}\\
{\small Batavia, IL 60510}\\
\\
{\sc Damien M. Pierce} \\ 
{\small Physics Department} \\
{\small Brookhaven National Laboratory} \\
{\small Upton, NY 11973}\\
}

\date{}
\maketitle
\thispagestyle{empty}

\begin{abstract}
\noindent
We determine the Tevatron's reach in supersymmetric parameter space in
trilepton, like-sign dilepton, and dilepton plus tau-jet channels. We
critically study the standard model background processes. We find
larger backgrounds and, hence, significantly smaller reach regions
than recent analyses. We identify the major cause of the background
discrepancy. We improve signal-to-noise by introducing an invariant
mass cut which takes advantage of a sharp edge in the signal dilepton
invariant mass distribution. Also, we independently vary the cuts at
each point in SUSY parameter space to determine the set which yields
the maximal reach. We find that this cut optimization can
significantly enhance the Tevatron reach.
\end{abstract}

\vspace{1.8cm}
\centerline{\sl Submitted to Phys. Rev. D}

\end{titlepage}

\setcounter{page}{1} 

\section{Introduction} \label{sec:introduction}

For at least the next 6 years the Fermilab Tevatron will remain the
highest energy collider in the world. The Tevatron upgrade will
provide an exciting opportunity for discovering physics beyond the
Standard Model.  The hadronic environment at the Tevatron presents a
number of challenges and extracting new physics signals can be
difficult. In this respect, signatures with low Standard Model
(especially QCD) backgrounds are extremely valuable, as they may
provide our best opportunity for finding new physics before the LHC
turns on.

Supersymmetry (SUSY) \cite{original} has been fascinating particle
physicists for more than 25 years.  It seems an intrinsic component of
theories unifying gravity and gauge interactions such as string
theory, M-theory or supergravity, and has played an important role in
the `second string revolution' of the last few years. The minimal
supersymmetric extension of the Standard Model (MSSM) is a
well-defined, renormalizable and calculable model, which offers a
technical solution to the hierarchy problem \cite{hierarchy}, if the
masses of the superpartners of the standard model particles are of
order the weak scale. Our belief that supersymmetry might be relevant
at energy scales accessible at present colliders is reinforced by the
successful gauge coupling unification \cite{gauge unification}. Also,
due to decoupling the MSSM is generally in agreement with precision
data \cite{precision fits}. In addition, a generic prediction of the
MSSM is the existence of a light Higgs boson \cite{light
Higgs, Higgs constraints}, which is preferred by fits to data
\cite{Higgs mass fit}.  In summary, the MSSM is a well-motivated
extension of the Standard Model, which has a very rich and interesting
phenomenology \cite{MSSM review}.

Because of the relatively low (compared to the LHC or NLC)
center of mass energy and integrated luminosity in Run II,
the Tevatron is able to explore only the low end of the
superpartner spectrum. Searches for colored superpartners
(squarks and gluinos) are done in jetty channels, which suffer
from relatively large backgrounds. On the other hand, $SU(2)$-gaugino
pair production leads to a unique clean trilepton signature
\cite{3L,Mrenna}, which has been considered a `gold-plated'
mode for SUSY discovery at the Tevatron. Both CDF and D0 have
already performed Run I trilepton analyses \cite{3L-exp}.
In light of the importance of this channel in Run II, it is clamant to
\begin{itemize}
\item have a reliable estimate of both signal and background rates.
We would prefer to determine background rates from data, but until Run
II we primarily rely on Monte Carlo simulations.  ISAJET and PYTHIA
have been two of the most commonly used event generators in SUSY
analyses.  While there is a reasonable agreement for the signal,
PYTHIA-based studies \cite{Mrenna,LM} have obtained larger values for
the trilepton backgrounds (mainly $WZ$ and $ZZ$) than ISAJET-based
analyses \cite{3L,BPT,BK}. This discrepancy was noticed and discussed
in the TeV 2000 Report \cite{TeV2000}, but was attributed to the
different lepton rapidity cuts used in the various analyses.
\item use an optimized set of cuts, which will maximize the Tevatron
reach. A first step in this direction was taken in Refs.~\cite{BK},
where softer lepton $p_T$ cuts have been proposed, thus enhancing
signal over background throughout a large part of parameter space.
\item include next-to-leading order (NLO) corrections to the
production cross sections. The corrections to diboson production
\cite{VV_corr}, $t\bar{t}$ production \cite{tt_corr} and Drell-Yan
\cite{DY_corr}, have been known for some time, and the corrections
will soon be available for chargino-neutralino production as well
\cite{Chi_corr}.
\item identify regions of parameter space where the reach via the
trilepton signature is diminished and try to find an alternative
search strategy in those regions. An example of this sort is the large
$\tan\beta$ region with light sleptons, where one often finds that
both the chargino and the neutralino decay predominantly to tau
leptons. Then the trilepton signal has a very small branching ratio
{\em and} the leptons are quite soft, which can make it unobservable
at Run II, even for chargino masses as low as 100 GeV. In this case it
is possible to recover sensitivity by considering alternative
signatures with tau jets \cite{LM}. Another alternative to the
trilepton signature is the inclusive like-sign dilepton channel
\cite{JN}, where the signal acceptance is increased by not requiring
the odd-sign lepton in the event.
\end{itemize}
In this paper, we shall try to address most of these issues.  We
perform detailed Monte Carlo simulations of signal and background
using both PYTHIA and ISAJET and explain the cause for the largest
background discrepancy. We also determine the maximum reach by
applying an optimal set of cuts at each point in the supersymmetric
parameter space.

Also, we make use of the presence of a sharp edge in the dilepton
invariant mass distribution of the signal by applying a more
restrictive invariant mass cut, thus reducing the $WZ$ and $ZZ$
backgrounds.  As the NLO corrections to gaugino production are not yet
available, we conservatively use leading order cross sections for all
processes.  Preliminary results \cite{Chi_corr} show that the
$k$-factor is roughly the same for both signal and background. Hence,
we expect that the Tevatron reach will be improved once NLO
corrections are incorporated.

We show our results for the discovery potential of the upgraded
Tevatron in the so-called minimal supergravity model (mSUGRA)
\cite{SUGRA}.  This model has universal soft parameter boundary
conditions at the grand unification scale, and its spectrum displays
characteristic properties. For example, the imposition of electroweak
symmetry breaking results in\footnote{$\mu$ is the Higgsino mass
parameter and $M_2$ is the soft supersymmetry breaking SU(2) gaugino
mass.}  $|\mu|>M_2$, so that the lightest chargino and lightest two
neutralinos are gaugino-like. Also, the squark and slepton masses are
generation independent, except at large $\tan\beta$ where the third
generation masses can be lighter. This model has five input
parameters: the scalar mass $M_0$, the gaugino mass $M_{1/2}$, the
$A$-term $A_0$, the ratio of vacuum expectation values $\tan\beta$,
and the sign of the $\mu$ term. We show results for $\mu>0$ and
$A_0=0$.

We adopt a signature driven approach by comparing and contrasting
three of the cleanest channels for Run II -- the trilepton (3L)
\cite{3L, Mrenna}, like-sign dilepton (2L) \cite{JN} and dilepton plus
tau jet (2L1T) \cite{LM} channels. The 3L channel is the long studied
``gold-plated'' channel. The 2L channel has larger signal acceptance
compared to 3L, but it is not {\it a priori} clear whether this
advantage will be spoiled by the concomitant increase in the
background. The 2L1T channel is known to be important at large
$\tan\beta$, where the right-handed tau-slepton is lighter than the
first two generation sleptons. Here we will discuss this channel at
small $\tan\beta$ as well.

We describe in detail our numerical analysis in
Section~\ref{sec:analysis}, where we also describe the cuts we
consider for each signature.  We discuss all non-negligible
backgrounds and their evaluation in Section~\ref{sec:backgrounds}.
Then, in Section~\ref{sec:SUGRA}, we map our results for the Tevatron
reach onto the parameter space of the mSUGRA model.
We reserve Section~\ref{sec:conclusions} for our conclusions.

\section{Analysis} \label{sec:analysis}

In this section we describe our numerical analysis.  We use
PYTHIA v. 6.115 \cite{PYTHIA} and ISAJET v. 7.42 \cite{ISAJET} for
event generation, and the SHW v. 2.2 detector simulation package
\cite{SHW}, which mimics an average of the CDF and D0 Run II detector
performance. We use PYTHIA for the background determinations, together
with TAUOLA \cite{tauola} to account for the correct (on average) tau
polarization in tau decays.  We have made several modifications in
SHW, which are appropriate for our purposes:
\begin{enumerate}
\item We extend the tracking coverage to $|\eta|<2.0$, which increases
the electron and muon acceptance, as is expected in Run II \cite{TDR}.
For muons with $1.5<|\eta|<2.0$, we apply the same fiducial efficiency
as for $1.0<|\eta|<1.5$. However,
we still require that tau jets are reconstructed only up to $|\eta|<1.5$.
\item We retain the existing electron isolation requirement and add a
muon isolation requirement $I<2$ GeV, where $I$ is the total
transverse energy contained in a cone of size $\Delta
R=\sqrt{\Delta\phi^2+\Delta\eta^2}=0.4$ around the muon.
\item We increase the jet cluster $E_T$ cut to 15 GeV and correct the
jet energy for muons. We also add a simple electron/photon rejection cut
$E_{em}/E_{had}<10$ to the jet reconstruction algorithm, where
$E_{em}$ ($E_{had}$) is the cluster energy from the
electromagnetic (hadronic) calorimeter.
\item We correct the calorimeter $\met$ for muons.
\item We account for an incorrect assignment of neutralino particle
id's in the ISAJET translation of STDHEP v. 4.05 \cite{STDHEP}
\footnote{The assignment has been corrected in STDHEP v. 4.06.}.
\end{enumerate}
The addition of the muon isolation cut and the jet $E_{em}/E_{had}$
cut allows us to uniquely resolve the ambiguity arising in SHW v. 2.2,
when a lepton and a jet are very close.

As we mentioned in the introduction, we show results for three of the
cleanest SUSY channels in Run II at the Tevatron -- trileptons,
inclusive like-sign dileptons and dileptons plus a tau jet. In our
analysis we consider both channel specific and channel independent
cuts. In most of those cases, we use several alternative values for
the cut on a particular variable. For example, we try several $\met$
cuts, several sets of $p_T$ cuts, etc. We employ a parameter space
dependent cut optimization: at each point in SUSY parameter space, we
consider all possible combinations of cuts, and determine the best
combination by maximizing $S/\sqrt{B}$.  In contrast to superior
neural network analyses, the additional CPU requirements when
employing this simple optimization are negligible.  We concede that it
may not be possible to perform an identical analysis with real data,
particularly due to trigger issues. Even so, it is useful and
interesting to see which cuts work best in the different parts of
parameter space, and to see how much one can gain by choosing optimal
cuts.

We first list the channel-independent cuts, which in general are designed
to suppress backgrounds common to all three channels.
\begin{enumerate}
\item Four $\met$ cuts: $\met>\{15,20,25\}$ GeV or no cut.
\item Six high-end invariant mass cuts for any pair of opposite sign,
same flavor leptons. The event is discarded if: $|M_Z -
m_{\ell^+\ell^-}|<\{10,15\}$ GeV; or $m_{\ell^+\ell^-}>\{50,60,70,80\}$ GeV.
\item Four azimuthal angle cuts on opposite sign, same flavor leptons:
two cuts on the difference of the azimuthal angle of the two highest
$p_T$ leptons, $|\Delta\varphi|<\{2.5,2.97\}$, one cut
$|\Delta\varphi|<2.5$ for {\em any pair} leptons, and no cut.
\item An optional jet veto (JV) on QCD jets in the event.
\end{enumerate}

We list the channel-specific $p_T$ cuts in Table~\ref{pt cuts}.  In
the 3L channel, the first four $p_T$ cuts in the table also require a
central lepton with $p_T>11$ GeV and $|\eta|<1.0$ or 1.5.

\begin{table}[t!]
\centering
\renewcommand{\arraystretch}{1.5}
\begin{tabular}{||c||c|c|c||}
\hline\hline
Channel   & \multicolumn{3}{c||}{$p_T$ cuts} \\
\hline\hline
3L   & $p_T(\ell_1)$ & $p_T(\ell_2)$ & $p_T(\ell_3)$ \\ \hline
1    &   11       &     5      &     5  \\ \hline
2    &   11       &     7      &     5  \\ \hline
3    &   11       &     7      &     7  \\ \hline
4    &   11       &    11      &    11  \\ \hline
5    &   20       &    15      &    10  \\ \hline
\hline
\end{tabular}
\vspace{5mm}
\begin{tabular}{||c||c|c||}
\hline\hline
2L   & $p_T(\ell_1)$ & $p_T(\ell_2)$ \\ \hline
\hline
1    &   11       &     9    \\ \hline
2    &   11       &    11    \\ \hline
3    &   13       &    13    \\ \hline
4    &   15       &    15    \\ \hline
5    &   20       &    20    \\ \hline
\hline
\end{tabular}
\vspace{5mm}
\begin{tabular}{||c||c|c|c||}
\hline
\hline
2L1T & $p_T(\ell_1)$ & $p_T(\ell_2)$ & $p_T(\tau)$\\ \hline
\hline
1    &    8       &     5      &    10   \\ \hline
2    &    8       &     5      &    15   \\ \hline
3    &   11       &     5      &    10   \\ \hline
4    &   11       &     5      &    15   \\ \hline
\hline
\end{tabular}
\parbox{5.5in}{
\caption{Channel-specific sets of $p_T$ cuts. \label{pt cuts}}}
\end{table}

For all channels we impose a low-end invariant mass cut on any pair of
opposite sign, same flavor leptons, $m_{\ell^+\ell^-}>11$ GeV. This
cut is designed to suppress a number of backgrounds, e.g. Drell-Yan,
$b\bar{b}$, $c\bar{c}$, and the contribution from $W\gamma^*$ (see
below). It is common for lepton analyses to include a cut on the
$(\Delta\eta,\Delta\varphi)$ distance between any two leptons $\Delta
R>0.4$, which suppresses background from $b\bar{b}$, $c\bar{c}$, and
anomalously reconstructed cosmics. We choose not to include this cut,
since we do not simulate those backgrounds. (Monte Carlo simulations
do not reliably estimate the backgrounds from $b\bar{b}$ and
$c\bar{c}$ production.) We have checked, however, that the effect of
the $\Delta R$ cut on signal and background is negligible.

In the next section we briefly discuss the main backgrounds for the
three channels. This will also motivate the choice of some of the cuts
above.

\section{Backgrounds}\label{sec:backgrounds}

We simulate the following background processes (with the generated
number of events in parentheses): $ZZ$ ($10^6$), $WZ$ ($10^6$), $WW$
($10^6$), $t\bar{t}$ ($10^6$), $Z+$jets ($8\cdot10^6$) and $W+$jets
($8\cdot10^6$).

\subsection{Backgrounds to the trilepton channel}

We start with the $WZ$ background, which is known to be the major
source of background for the 3L channel. The total $WZ$ cross section
at Run II will be $\sim 2.6$ pb. Folding in the branching ratios of
$W$ and $Z$ to leptons, we get a 3L WZ background cross section
production of 46 fb.  It has a reducible and an irreducible
component. The irreducible component ($\sim 3$ fb) is due to
$Z\rightarrow\tau^+\tau^-\rightarrow \ell^+\ell^-$ decays. The
invariant mass of the resulting lepton pair from the tau decays is
usually far from the $Z$-mass, in a region which is typical of the
signal.  Hence there is no obvious cut which can substantially reduce
this part of the $WZ$ background without at the same time reducing
signal-to-noise. On the other hand, the remaining background ($\sim
43$ fb) is reducible, since it arises from $Z\rightarrow \ell^+\ell^-$
decays.  In this case the invariant mass $m_{\ell^+\ell^-}$ of the
resulting lepton pair is equal to $p_Z^2 = (p_Z)^\mu (p_Z)_\mu$, where
$(p_Z)_\mu$ is the 4-momentum of the parent boson.  Most of the time
the $Z$ is produced nearly on-shell, $p_Z^2 \approx M_Z^2$. Hence, the
invariant mass cut $|m_{\ell^+\ell^-}-M_Z|>10$ GeV is very efficient
in removing this source of background. However, the parent can also be
off-shell, due to either the $Z$-width, or to $WZ$/$W\gamma$
interference.  In PYTHIA, where only the first effect is modeled, the
lepton pair invariant mass distribution follows a Breit-Wigner
shape. We find that roughly 10\% of the 3L background events pass the
dilepton invariant mass cut, thus bringing the reducible background
cross section down to about 4.3 fb. This is almost a factor of two
larger than the corresponding irreducible background cross section
(compare to 2.6 fb). Since ISAJET does not incorporate either $Z$-tail
effect, we find essentially no reducible background cross section from
ISAJET after the dilepton invariant mass cut is
applied\footnote{Energy smearing in the detector simulation produces a
very small background which survives the 10 GeV $Z$-mass window cut.}.
This difference between ISAJET and PYTHIA largely accounts for
the discrepancy in the backgrounds found in Refs.~\cite{Mrenna,LM} and
Refs.~\cite{3L,BPT,BK}. In what follows we use PYTHIA for our
background estimate.

\begin{figure}[t]
\centerline{\psfig{figure=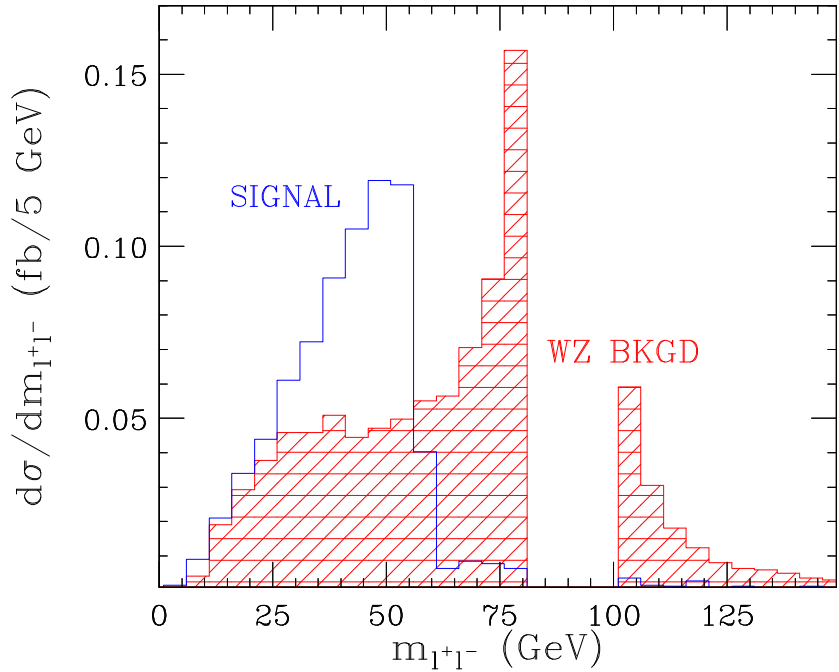,height=3.5in}}
\begin{center}
\parbox{5.5in}{
\caption[] {\small The invariant mass distribution of any pair
opposite sign, same flavor leptons for the signal events (with
$M_0=700$ GeV, $M_{1/2}=160$ GeV, $\tan\beta=5$) and the PYTHIA $WZ$
background. We impose a set of cuts from Ref.~\cite{BK}:
$p_T(\ell)>\{11,7,5\}$ GeV, central lepton with $p_T>11$ GeV and
$|\eta|<1.0$, $\met>25$ GeV and $|m_{\ell^+\ell^-}-M_Z|>10$ GeV.  Each
histogram is normalized to its cross section.
\label{inv mass}}}
\end{center}
\end{figure}

We illustrate the above discussion in Fig.~\ref{inv mass}
where we show the invariant mass distribution of {\em any} pair of
opposite sign, same flavor leptons for both signal and $WZ$
background.  The signal point has $M_0=700$ GeV, $M_{1/2}=160$ GeV and
$\tan\beta=5$, which results in $m_{\tilde\chi_1^{\pm}}\simeq
m_{\tilde\chi_2^0}\simeq122$ GeV. The leptons are required to pass the
set of cuts from Ref.~\cite{BK}: $p_T(\ell)>\{11,7,5\}$ GeV, central
lepton with $p_T>11$ GeV and $|\eta|<1.0$, $\met>25$ GeV and
$|m_{\ell^+\ell^-}-M_Z|>10$ GeV.  The histograms are normalized to the
respective cross section.  

Even with the modeling of the $Z$-width effect, we caution that the
$WZ$ simulation in PYTHIA is still not realistic, since the
$W\gamma^*$ contribution is neglected. It results in a peak at low
invariant mass, and the resulting distribution is markedly different
from the result shown in Fig.~\ref{inv mass}. We anticipate that it
will be necessary to cut away {\em all} events on the low-end of the
dilepton invariant mass distribution. We therefore always apply the
cut $|m_{\ell^+\ell^-}|>11$ GeV for all three channels. This cut also
helps in eliminating background lepton pairs from Drell-Yan, as well
as $J/\psi$ and $\Upsilon$ decays.

Since recent trilepton analyses of the Tevatron reach \cite{BPT,BK}
use ISAJET for the simulation, they underestimate the trilepton
background.  As a result, we find a significantly reduced Tevatron
reach. However, in some cases we can employ an invariant mass cut
which will substantially reduce the $WZ$ background with no
significant loss of signal.  The signal distribution in Fig.~\ref{inv
mass} has a sharp kinematic cut-off at around
$m_{\tilde\chi_2^0}-m_{\tilde\chi_1^0}\sim
m_{\tilde\chi^{\pm}_1}/2\sim 60$ GeV, and we can exploit this feature
to increase signal-to-noise. Indeed, by applying the more stringent
cut $m_{\ell^+\ell^-}>m_{\tilde\chi^{\pm}_1}/2 \sim 60$ GeV, we can
eliminate most of the off-shell $Z$ events, at almost no cost to
signal. This is why in addition to standard $Z$-mass window cuts we
consider four dilepton mass cuts which eliminate {\em all} events
above a given invariant mass value. Also notice that only a very small
fraction of signal events have invariant dilepton masses between 0 and
11 GeV, so that the low-end invariant mass cut $m_{\ell^+\ell^-}>11$
GeV is quite efficient in improving $S/\sqrt{B}$.  Our discussion of
the $WZ$ background can be similarly applied to the less serious, but
nevertheless non-negligible $ZZ$ background.

The second largest background to the 3L channel is from dilepton
$t\bar{t}$ events, where there happens to be a third isolated lepton
from a $b$-jet. This background is most easily suppressed by a jet
veto.

Finally, the remaining 3L background one should worry about is $Z$-jet
production. In this case part of the background is due to events where
a jet fakes a lepton. Monte Carlo simulations, especially with a
simplified detector simulation like SHW, cannot give a reliable
estimate of this background.  In order to estimate the fake rate one
has to understand the details of the detector response as well as the
jet fragmentation. Only with Run II data will one be able to obtain a
good estimate. For our study we follow a procedure which makes use of
Run I data. It was used in Ref.~\cite{JN} to study the $W+{\rm jets}$
background to the 2L channel (see the next subsection).

\subsection{Backgrounds to the like-sign dilepton channel}\label{sec:2L}

We now discuss the backgrounds to the 2L channel.  Ref.~\cite{JN}
observed that it can be advantageous to not require the odd-sign
lepton in the 3L events due to the gain in signal acceptance. At the
same time, events with two like-sign leptons are still quite rare at
the Tevatron, so the 2L channel was suggested as a possible
alternative to 3L for SUSY searches in Run I. However, the rates for
the relevant diboson backgrounds were somewhat underestimated, since
ISAJET was used for the simulation.  Here we are interested in
determining whether the 2L channel will be useful in the larger
background environment of Run II.

We first do a back-of-the-envelope comparison of the $WZ$ backgrounds
for the 3L and 2L channels. After not requiring the odd-sign lepton,
one is left with the choice of {\em vetoing} that lepton. In the case
of a veto, we find for the relative size of the two backgrounds
\beq
{\sigma_{WZ}({\rm 2L}) \over \sigma_{WZ}({\rm 3L})}\simeq
{2\varepsilon_l (1-\varepsilon_l) +0.35\varepsilon_\tau \left[
0.65+0.35(1-\varepsilon_\tau) \right] \over
2\varepsilon_l^2\varepsilon_Z + 0.35^2\varepsilon^2_\tau} \sim
{1-\varepsilon_l \over\varepsilon_l\varepsilon_Z},
\label{2L/3L veto}
\eeq 
where $\varepsilon_l$ is the acceptance for leptons coming directly
from $W$ or $Z$ decays, $\varepsilon_\tau$ is the acceptance for the
(usually softer) leptons coming from leptonic tau decays, and
$\varepsilon_Z$ is the efficiency of the $Z$-window invariant mass
cut: $|m_{\ell^+\ell^-}-M_Z|>10$ GeV (we have neglected its effect on
the {\em irreducible} background component). We find from Monte Carlo
that the typical Run II values for these efficiencies in $WZ$
production are $\varepsilon_l\sim 0.60$, $\varepsilon_\tau\sim 0.46$
and $\varepsilon_Z\sim 0.10$.  Plugging into Eq.~(\ref{2L/3L veto}),
we find for the ratio 6.3, which agrees reasonably well with the
result 5.8 from our full Monte Carlo simulation. For the signal point
shown in Fig.~\ref{inv mass} we find $\varepsilon_l=0.62$ and
$\varepsilon_Z=0.99$ and the corresponding ratio is \beq
{\sigma_{signal}({\rm 2L}) \over \sigma_{signal}({\rm 3L})}\simeq
{1-\varepsilon_l \over\varepsilon_l\varepsilon_Z}\sim 0.6.  \eeq This
reveals that vetoing the third lepton is definitely not a good idea.
In comparison to the 3L channel, the signal goes down, while the major
background component is increased almost 6 times!

We therefore only consider the {\em inclusive} 2L channel, where we do
not have any requirements on the third (odd-sign) lepton, just as in
Ref.~\cite{JN}. In this case, the signal acceptance is definitely
increased. Unfortunately, the corresponding increase in the background
is even larger than before: \beq {\sigma_{WZ}({\rm 2L}) \over
\sigma_{WZ}({\rm 3L})}\simeq {2\varepsilon_l
(1-\varepsilon_l+\varepsilon_l\varepsilon_Z) +0.35\varepsilon_\tau
\over 2\varepsilon_l^2\varepsilon_Z + 0.35^2\varepsilon^2_\tau} \sim
7.3
\label{2L/3L}
\eeq for the typical values of the efficiencies. We can see
immediately that the 2L channel can compete with the 3L on the basis
of $S/\sqrt{B}$ only if the lepton acceptance for the signal is less
than $1/\sqrt{7.3}\sim 37\%$. However, for typical values of the SUSY
model parameters the lepton acceptance is much higher.

\begin{figure}[t]
\centerline{\psfig{figure=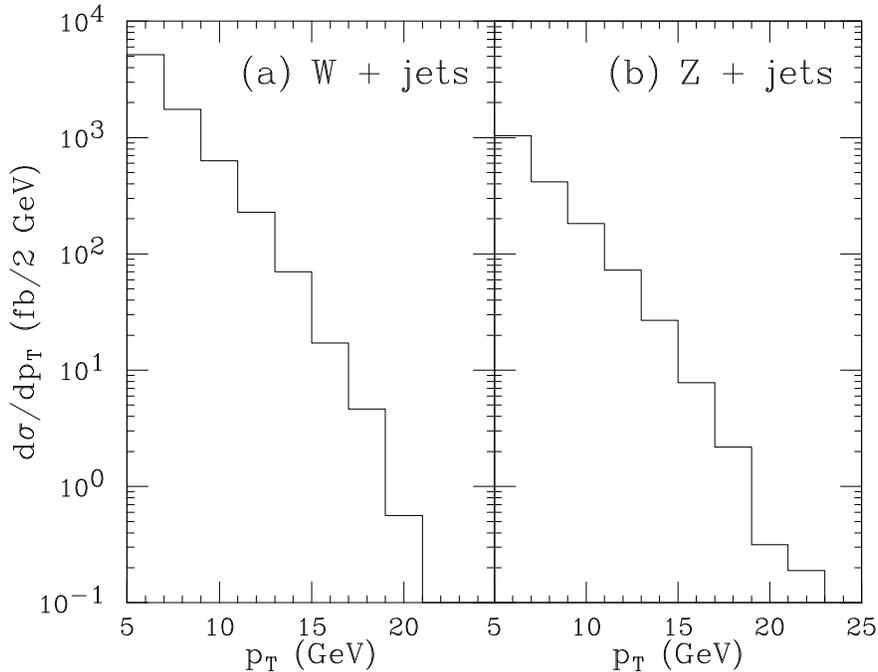,height=3.5in}}
\begin{center}
\parbox{5.5in}{
\caption[] {\small $p_T$ distribution of isolated tracks in (a)
$W+$jet production and (b) $Z+$jet production. The isolated tracks
have $I<2$ GeV and are outside the $\Delta R=0.7$ cone around any
jet. The events in the distributions potentially contribute to the 2L
background: they have one real lepton with $p_T>11$ and $|\eta|<2$ and
one same sign isolated track with $|\eta|<2$.
\label{pT_iso}}}
\end{center}
\end{figure}

To make matters worse, the 2L channel suffers from a potentially large
new source of background: $W+$jet production where the jet fakes a
lepton.  Although the rate for a jet faking a lepton is quite small,
on the order of $10^{-4}$, the large $W+$jet cross section results in
a major background for the 2L channel. As we mentioned earlier, the
best way to estimate this background is from data, since Monte Carlo
simulations are not reliable for fakes. In our analysis we shall
follow the procedure of Ref.~\cite{JN}, where the rate for observing
an isolated track which would otherwise pass the lepton cuts was
measured in the Run I $Z+$jet event sample.  This rate was then
multiplied by the probability that, given an isolated track, it would
fake a lepton. This probability was measured in Run I minimum bias
events to be $\sim 1.5\%$, independent of $p_T$ \cite{JN}. In our
study we first simulate with Monte Carlo the $p_T$ distribution of
isolated tracks in $W$ and $Z$ production, which is shown in
Fig.~\ref{pT_iso}. In this figure we plot the isolated track $p_T$
distribution corresponding to the 2L background, i.e. we combine a
real lepton with $p_T>11$ GeV and $|\eta|<2$ with a same sign isolated
track with $|\eta|<2$. Hence, the 2L background cross section is
obtained by multiplying the cross section from Fig.~\ref{pT_iso} by
the probability that an isolated track will fake a lepton. We see that
the background from fakes falls extremely fast with $p_T$, so a larger
$p_T$ requirement will substantially suppress it.

We normalize the isolated track rate to data. Using the measured 1.5\%
fake rate per isolated track, we find 1.5 fb of cross section when
running the simulation at $\sqrt{s}=1800$ GeV and using the set of
cuts from Ref.~\cite{JN}. This is half the cross section found in
Ref.~\cite{JN}. Hence, to match the data to PYTHIA/SHW we need to
double the isolated track rate obtained from Monte Carlo.

\subsection{Backgrounds to the dilepton plus tau jet channel}

The largest background to the 2L1T channel is Drell-Yan \cite{LM},
where the tau jet is a fake. As it turns out, SHW does quite a good
job in simulating the fake tau rate \cite{fake taus}, so we can safely
rely on the Monte Carlo for this background.  We shall not further
discuss the backgrounds to the 2L1T channel; we refer the interested
reader to Refs.~\cite{LM,Workshop report}.  We find a marginal increase
in the background rate here, due to the larger $\eta$ coverage used in
this analysis.

\subsection{Summary and discussion}\label{sec:summary}

In Table~\ref{Table_BG} we summarize our results for the
different backgrounds to the three channels.
\begin{table}[t!]
\centering
\renewcommand{\arraystretch}{1.5}
\begin{tabular}{||c||c|c||c|c||c||}
\hline\hline
Process    &       3L         &  \cite{BK}  
             &        2L        &  \cite{JN} 
               &        2L1T        \\ \hline
           &                  &    
             & $E=2$ TeV      &  $E=1.8$ TeV
               &                 \\  
           &                  &    
             & $|\eta(\ell)|<2$  &  $|\eta(\ell)|<1$
               &                 \\  \hline\hline      
$ZZ$       & 0.21  $\pm$ 0.01 & 0.04 
             & 1.836 $\pm$ 0.006 & 0.1 $\pm$ 0.01
               & 0.372 $\pm$ 0.003                  \\ \hline
$WZ$       & 1.39  $\pm$ 0.01 & 0.40 
             & 8.79 $\pm$ 0.03 & 1.1 $\pm$ 0.02
               & 1.36  $\pm$ 0.01                   \\ \hline
$WW$       & 0.009 $\pm$ 0.003& ---  
             & 0.002 $\pm$ 0.001 & 0   $+$ 0.02
               & 0.54  $\pm$ 0.02                   \\ \hline
$t\bar{t}$ & 0.33  $\pm$ 0.01 & 0.14 
             & 0.21 $\pm$ 0.01 & 0   $+$ 0.02
               & 1.64  $\pm$ 0.03                   \\ \hline
$Z+$jet     & 0.13 $\pm$ 0.01  & ---  
             & 3.58 $\pm$ 0.05 & 1.1 $\pm$ 0.1
               & 11.3 $\pm$ 0.6                         \\ \hline
$W+$jet     &        ---       & ---  
             & 10.5 $\pm$ 0.2  & 3.0
               & ---                         \\ \hline\hline
total      & 2.07 $\pm$ 0.02  & 0.58 
             & 24.9 $\pm$ 0.2    & 5.6
               & 15.2 $\pm$ 0.6                         \\ \hline\hline
\end{tabular}
\parbox{5.5in}{
\caption{ Background cross sections after cuts (in fb) for the three
channels, each with a set of cuts described in the text.  All errors
are statistical.  We also list the 3L background found in
Ref.~\cite{BK} and the 2L background found in Ref.~\cite{JN}.
\label{Table_BG}}}
\end{table}

The results are presented for a standard choice of cuts in each case.
For the 3L channel, we pick the set of cuts from Ref.~\cite{BK}:
$p_T(\ell)>\{11,7,5\}$ GeV, central lepton with $p_T>11$ GeV and
$|\eta|<1.0$, $\met>25$ GeV, $|m_{\ell^+\ell^-}-M_Z|>10$ GeV, and no
$|\Delta\varphi|$ or JV cuts.  For the 2L channel we use the following
cuts from Ref.~\cite{JN}: $p_T(\ell)>\{11,11\}$ GeV,
$|m_{\ell^+\ell^-}-M_Z|>10$ GeV, and no $\met,\ |\Delta\varphi|$ or JV
cuts. Here we require both leptons to have $|\eta|<2.0$, whereas
Ref.~\cite{JN} requires $|\eta|\lsim 1$. Finally, for the 2L1T channel
we use the cuts $p_T(\tau)>15$ GeV, $|\eta(\tau)|<1.5$,
$p_T(\ell)>\{8,5\}$ GeV, $|\eta(\ell)|<2.0$,
$|m_{\ell^+\ell^-}-M_Z|>10$ GeV, $\met>20$ GeV, and no JV, from
Ref.~\cite{LM}. All errors in the table are statistical.

Comparing to Refs.~\cite{BK,JN} we see a significant increase in the
$WZ$ and $ZZ$ backgrounds to both the 3L and 2L channels. In the 2L
case this is in part due to the larger $\eta$ coverage that we use, as
well as the higher center-of-mass energy. The remaining difference is
due to the lack of $Z$-tail effects in the ISAJET simulation. The $WZ$
background is almost as large as $W+$jet, which is not surprising
based on the estimate in Eq.~(\ref{2L/3L}) (compare the $WZ$
backgrounds to 3L versus 2L).  We remind the reader that our 2L
$W+$jet background has been normalized to the 2L $W+$jet background
from Ref.~\cite{JN}, so the difference in the $W+$jet background
seen in the table is due solely to the different tracking coverage
and center of mass energy.

In the 3L case, we find significantly larger backgrounds than
Ref.~\cite{BK}. In fact, we surmise that the lepton efficiency is
smaller in SHW than in the ISAJET detector simulation, and if one were
to take this into account the background differences would be even
larger. Part of the difference in the $WZ$ and $ZZ$ backgrounds is due
to the fact that PYTHIA gives a $\sim15\%$ larger diboson cross
section than ISAJET. However, the largest part of the difference in
the diboson background rates is due to the $Z$-width. Notice that with
the fake rate procedure discussed in Sec.~\ref{sec:2L} we are able to
obtain an estimate of the $Z+$jet trilepton background (where the jet
fakes a lepton). This background has not been taken into account in
previous studies.

We do not trust the Monte Carlo simulation to provide a reliable
estimate for the $W+$jet and $Z+$jet backgrounds where the jet gives
rise to a real isolated lepton. We expect these backgrounds to be
small, and we ignore them.

In the next section, we present results for the Tevatron reach in the
three channels in the minimal gravity-mediated SUSY breaking model.
In our simulations we use PYTHIA for the background estimates and
ISAJET for the signal.

\section{Discovery reach for gravity-mediated models} \label{sec:SUGRA}

We next discuss the discovery potential of the upgraded Tevatron in
the mSUGRA model. There are various SUSY production processes which
give rise to the lepton signals under consideration. The dominant
source of signal in most regions of parameter space is
$\tilde\chi_1^+\tilde\chi_2^0$ production. Other chargino/neutralino
and slepton production processes also contribute. Typically these
processes constitute a small fraction of the total signal cross
section, but in some regions of parameter space (e.g. large
$\tan\beta$, small $M_0$) they can dominate. We ignore the possibility
of small contributions from squark and gluino production processes.

We first review which parts of the mSUGRA parameter space are
accessible in Run II via the 3L signature. The
$\tilde\chi_1^+\tilde\chi_2^0$ production cross section depends
primarily on the chargino mass, and scales roughly as
$M_{1/2}^{-5.5}$. Therefore, at the Tevatron we can only explore
regions with small $M_{1/2}$, where the cross section is large. The
$\tilde\chi_1^+\tilde\chi_2^0$ production proceeds predominantly via
$s$-channel $W$-boson exchange. There is also destructive interference
from $t$-channel squark exchange. As a result, the
$\tilde\chi_1^+\tilde\chi_2^0$ cross section is slightly enhanced in
case of heavy squarks (i.e. at large values of $M_0$).

\begin{figure}[t]
\centerline{\psfig{figure=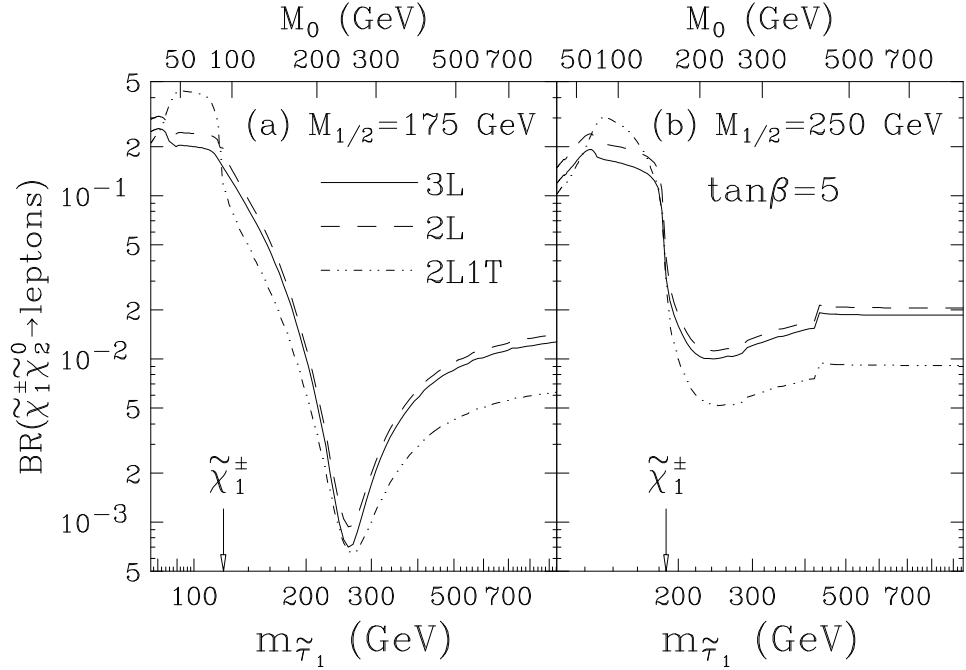,height=3.5in}}
\begin{center}
\parbox{5.5in}{
\caption[] {\small Branching ratios of a chargino-neutralino pair into
the 3L (solid), 2L (dash) and 2L1T (dot dot dash) channels versus the
lightest stau mass (bottom axis), or alternatively, versus $M_0$ (top
axis), with mSUGRA model parameters $\tan\beta=5$, $\mu>0$, $A_0=0$
and (a) $M_{1/2}=175$ GeV or (b) $M_{1/2}=250$ GeV. The arrows
indicate the chargino threshold $m_{\tilde\chi_1^\pm}=m_{\tilde
\tau_1}$. For the range of $M_0$ values shown the chargino mass
varies from 118 to 139 GeV in (a), and from 187 to 201 GeV in (b).
\label{br_frac_lowtb}}}
\end{center}
\end{figure}

At large values of $M_0$ ($M_0\gsim700$ GeV) the sleptons are heavy
and the gauginos decay to three-body final states dominantly via real
(at large $M_{1/2}$) or virtual (small $M_{1/2}$) $W$- or $Z$-boson
exchange. In this case the reach is determined solely by the signal
production cross section.  At smaller values of $M_0$ the off-shell
slepton mediated decays destructively interfere with the gauge
mediated decays. At small values of $M_0$ ($M_0\lsim M_{1/2})$
sleptons become lighter than the $\tilde\chi_1^+/\tilde\chi_2^0$ and
two-body decays to lepton final states open up. This enhances the
leptonic branching ratios of the gauginos and improves the Tevatron
reach. We illustrate these features in Fig.~\ref{br_frac_lowtb} where
we show the branching ratios\footnote{When we use the term `branching
ratios' in relation to a pair of particles, we mean a sum of products
of branching ratios. For example, if the two-body decays of
$\tilde\chi_1^+$ and $\tilde\chi_2^0$ are closed,
BR$(\tilde\chi_1^+\tilde\chi_2^0 \rightarrow 3L) \equiv \left[{\rm BR}
(\tilde \chi_1^+ \rightarrow \tilde \chi_1^0\ell^+\nu_\ell) + {\rm BR}
(\tilde \chi_1^+ \rightarrow \tilde \chi_1^0 \tau^ + \nu_\tau
\rightarrow \tilde\chi_1^0\ell^+ \nu_\ell\bar \nu_\tau \nu_\tau
)\right]\left[{\rm BR} (\tilde \chi_2^0 \rightarrow \tilde
\chi_1^0\ell^+\ell^-) + {\rm BR} (\tilde \chi_2^0 \rightarrow \tilde
\chi_1^0\tau^+\tau^- \rightarrow\right.$$ \tilde \chi_1^0 $
$\ell^+ \ell^- \bar\nu_\ell$ $\nu_\ell\bar\nu_\tau$
$\left.\nu_\tau)\right]$.}  of a chargino-neutralino pair into the 3L,
2L and 2L1T channels at an mSUGRA model point with $\tan\beta=5$,
$\mu>0$, $A_0=0$ and (a) $M_{1/2}=175$ GeV or (b) $M_{1/2}=250$ GeV.
We plot versus $m_{\tilde \tau_1}$ (bottom axis) or $M_0$ (top axis).
Notice that because of the rather small value of $\tan\beta$, all
slepton flavors are practically degenerate.

In Fig.~\ref{br_frac_lowtb}(a) the chargino mass
$m_{\tilde\chi_1^\pm}$ is only about 120 GeV, so the two-body decay to
the $W$-boson is closed. If $M_0>92$ GeV the chargino decays are
three-body and the 2L and 3L channels have a larger branching ratio
than the 2L1T, roughly by a factor of two.  The dip in the leptonic
branching ratios near $m_{\tilde \tau_1}\sim 260$ GeV is due to
destructive interference between the $Z$ and $\tilde\ell$-mediated
graphs \cite{BR interference}. In the region $M_0<92$ GeV the two-body
decays to sleptons are open and quickly become dominant. But notice
that while $\tilde\chi_2^0$ can decay to all slepton flavors through
its bino component, $\tilde\chi_1^\pm$ decays dominantly to a stau,
since the chargino couplings to the right-handed sleptons are
proportional to the corresponding lepton masses. Hence, the branching
fraction of the 2L1T channel dominates the region $40\lsim M_0\lsim90$
GeV. For $M_0\lsim40$ GeV the 2L1T branching ratio rapidly drops while
the 2L and 3L branching ratios increase and become dominant again, due
to two-body decays of $\tilde\chi^\pm_1$ to sneutrinos.

In Fig.~\ref{br_frac_lowtb}(b) the mass of $\tilde\chi_1^\pm$ is near
190 GeV and it is always above the $W$-boson threshold. If $M_0$ is
greater than about 400 GeV the $\tilde\chi_2^0$ is above the $Z$-boson
threshold, and in this region the branching ratios for the three
signatures are determined by the $W$- and $Z$-boson branching ratios
to leptons. In the region $M_0<150$ GeV the sleptons are lighter than
the chargino, and the two-body decay modes to sleptons greatly enhance
the leptonic branching fractions. The 2L1T channel does not become as
dominant because the chargino decays to quarks via an on-shell
$W$-boson.  The decrease in the leptonic branching ratios below
$M_0\simeq70$ GeV is as before due to two-body decays to sneutrinos.

\begin{figure}[t]
\centerline{\psfig{figure=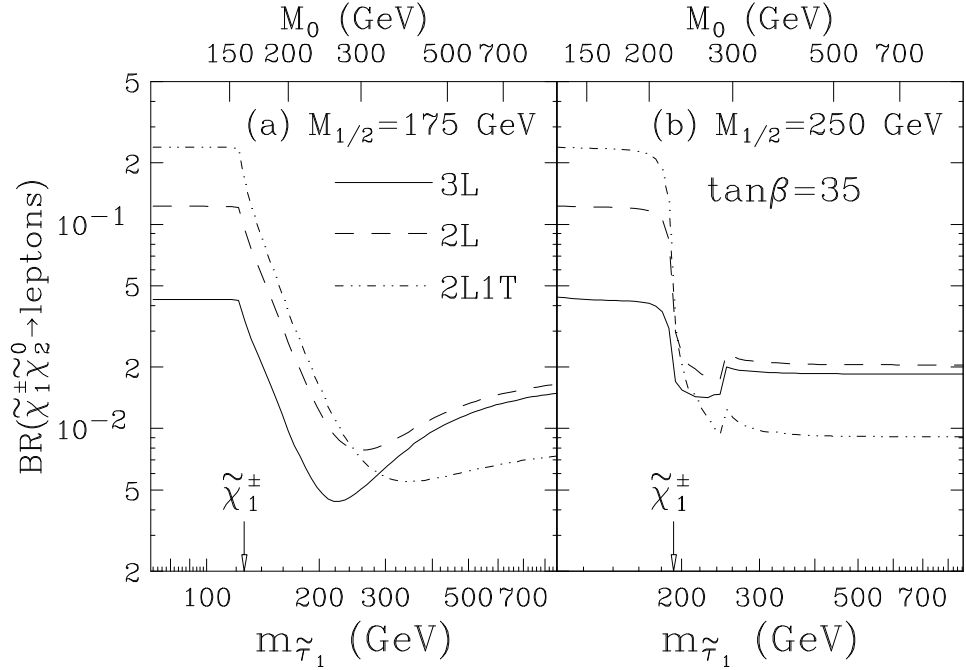,height=3.5in}}
\begin{center}
\parbox{5.5in}{
\caption[] {\small The same as Fig.~\ref{br_frac_lowtb}, but for
$\tan\beta=35$. This time the chargino mass varies from
126 to 140 GeV (in (a)), and from 192 to 201 GeV (in (b)).
\label{br_frac_hightb}}}
\end{center}
\end{figure}

In Fig.~\ref{br_frac_hightb} we show similar plots of the
branching fractions, but for $\tan\beta=35$. 
Many of the features seen in Fig.~\ref{br_frac_lowtb} are present here
as well. For example, in Fig.~\ref{br_frac_hightb}(a) we see the broad
region ($150\lsim m_{\tilde \tau_1}\lsim 400$ GeV) where the
destructive $Z$-$\tilde\ell$ interference severely diminishes the
branching ratios. And again we observe the $\tilde\chi_2^0\rightarrow
\tilde\chi_1^0 Z$ threshold in Fig.~\ref{br_frac_hightb}(b), near
$m_{\tilde \tau_1}=260$ GeV.  However, in this case the tau slepton is
significantly lighter than the first two generation sleptons. Hence,
the branching ratio to the three tau final state quickly approaches
100\% below the stau threshold, and the 2L1T branching ratio is large.
We also see that the 2L channel is competitive at small values of
$M_0$, and is preferred over 3L on the basis of branching ratio by a
factor of 2.8.

To summarize, we see that depending on the values of the mSUGRA
parameters, any of the three channels offers some promise to be
observed in Run II. In the rest of this section, we shall do a
comparative study of the three channels, accounting for all relevant
background processes and using a realistic detector simulation. We
shall not only update the existing 3L analyses with improved estimates
of the $WZ$, $ZZ$ and Drell-Yan background rates, but foremost we are
interested in evaluating the different prospects each channel can
offer. For example, we would like to see whether the 2L channel offers
reach beyond the 3L channel, or can cover regions where the 3L channel
is suppressed. Also, we want to determine the region of parameter
space where the 2L1T channel can offer an independent check of a
signal in one of the other channels.

The cut optimization procedure is an important ingredient in our
analysis. We show results for the reach with the optimal set of cuts,
determined independently at each point in SUSY parameter space. The
optimal set of cuts maximizes $S/\sqrt{B}$. We require the observation
of at least 5 signal events, and present our results as $3\sigma$
exclusion contours in the $M_0-M_{1/2}$ plane, for two representative
values of $\tan\beta$ -- 5 and 35. We fix $\mu>0$ and $A_0=0$.

\begin{figure}[t]
\centerline{\psfig{figure=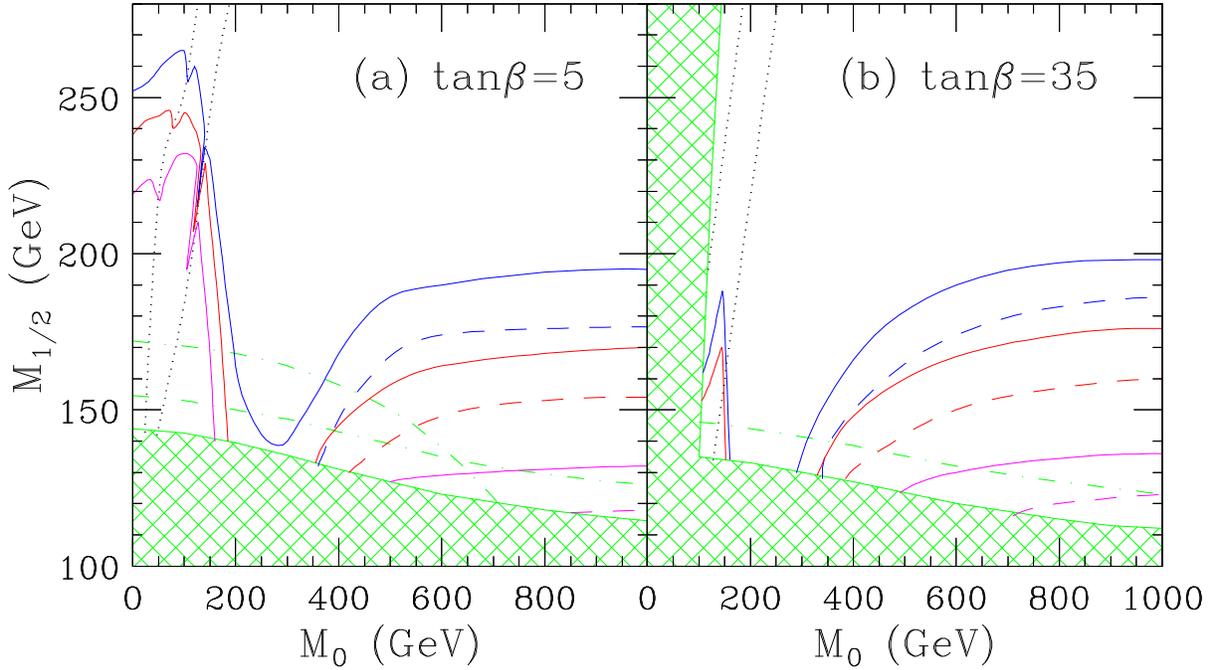,height=3.5in}}
\begin{center}
\parbox{5.5in}{
\caption[] {\small Tevatron reach in the 3L channel for mSUGRA models
with $\mu>0$, $A_0=0$, and (a) $\tan\beta=5$ or (b) $\tan\beta=35$. We
show the reach with both a standard set of soft cuts \cite{BK}
(dashed, for large $M_0$ only), as well as with the optimal set of
cuts (solid) (see text).  The reach is shown for 30 ${\rm fb}^{-1}$,
10 ${\rm fb}^{-1}$ and 2 ${\rm fb}^{-1}$ total integrated luminosity
(from top to bottom).  The cross-hatched region is excluded by current
limits on the superpartner masses. The dot-dashed lines correspond to
the projected LEP-II reach for the chargino and the lightest Higgs
masses. In Fig.~(a) the left dotted line shows where
$m_{\tilde\nu_\tau}=m_{\tilde\chi_1^\pm}$ and the right dotted line
indicates $m_{\tilde\tau_1}=m_{\tilde\chi_1^\pm}$. In Fig.~(b) the
dotted lines show where $m_{\tilde e_R}=m_{\tilde\chi_1^\pm}$ (left)
and $m_{\tilde\tau_1}=m_{\tilde\chi_1^\pm}$ (right).
\label{m0mh_3L}}}
\end{center}
\end{figure}

In Fig.~\ref{m0mh_3L} we show the Tevatron reach in the 3L channel
with a standard set of soft cuts \cite{BK} (dashed lines, for large
$M_0$ only to prevent crowding), as well as with the optimal set of
cuts (solid). We show the expected reach for 2, 10 and 30 ${\rm
fb}^{-1}$ total integrated luminosity.  The cross-hatched region is
excluded by current limits on the superpartner masses and the
dot-dashed lines indicate the projected LEP-II reach for the chargino
and the lightest Higgs mass\footnote{It should be kept in mind that at
small $\tan\beta$ the Higgs mass is a sensitive function of
$\tan\beta$. For example, if the Higgs-boson is not discovered at
LEP-II, $\tan\beta=3$ will be excluded. The reach contours are much
less sensitive to $\tan\beta$.}.
In Fig.~\ref{m0mh_3L}(a) the left (right) dotted line marks the
$\tilde\nu_\tau/\tilde\chi_1^\pm$ ($\tilde\tau_1/\tilde\chi_1^\pm$)
mass threshold, while in Fig.~\ref{m0mh_3L}(b) the left (right) dotted
line marks the $\tilde e_R/\tilde\chi_1^\pm$
($\tilde\tau_1/\tilde\chi_1^\pm$) mass threshold.

Comparing to the results of Refs.~\cite{BPT,BK}, we see the Tevatron
reach indicated in Fig.~\ref{m0mh_3L} is greatly reduced, due to the
larger backgrounds found in our analysis. Even with optimized cuts our
results show a much smaller observable region. For example, using the
set of cuts of Ref.~\cite{BK} at $\tan\beta=35$ and $M_0=1$ TeV, our
results indicate that the region bounded by the 2 fb$^{-1}$ 3-$\sigma$
contour extends to $M_{1/2}=123$ GeV (or 136 GeV with optimal cuts),
whereas in Ref.~\cite{BK} the corresponding 5-$\sigma$ region extends
to 180 GeV. Similarly, their 30 fb$^{-1}$ 3-$\sigma$ contour extends
to $M_{1/2}=250$ GeV, while ours extends to $M_{1/2}=186$ GeV (198 GeV
with optimal cuts).

\begin{figure}[t]
\centerline{\psfig{figure=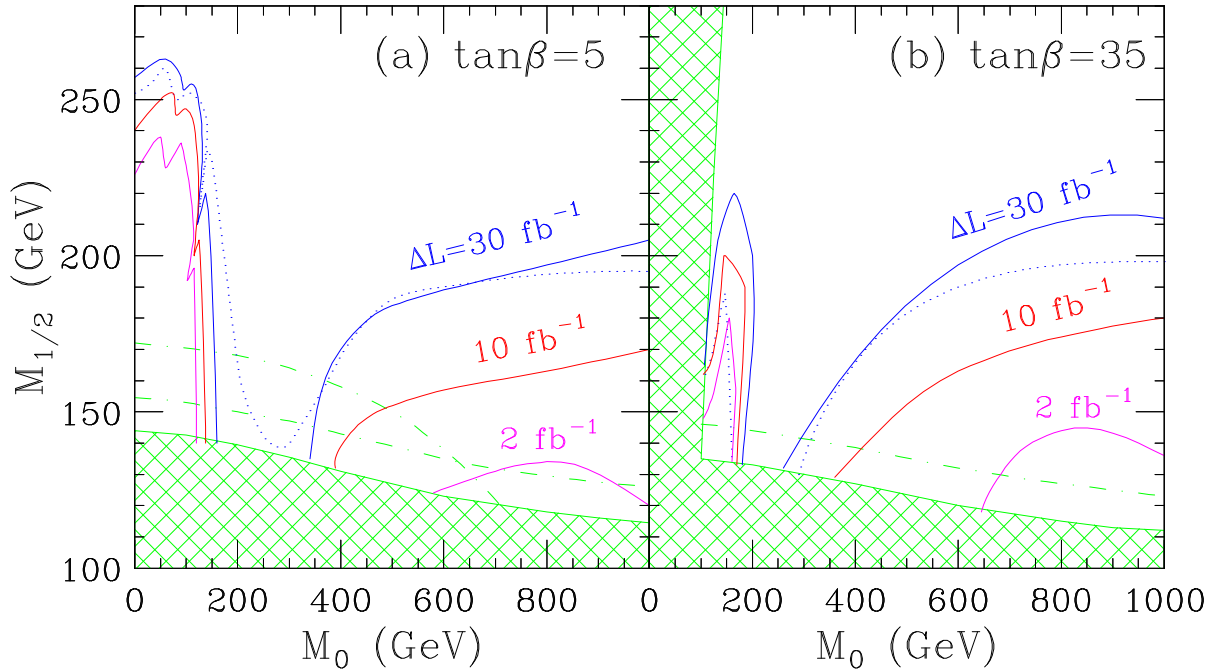,height=3.5in}}
\begin{center}
\parbox{5.5in}{
\caption[] {\small The difference $\Delta L$ (in ${\rm fb}^{-1}$)
between the required total integrated luminosity for the optimal set
of cuts and the default set of cuts, for the 3L signal.  The three
(solid) contours correspond to (from top to bottom) $\Delta L = 30,
10$ and 2 ${\rm fb}^{-1}$.  The dotted lines indicate the optimal-cut
30 ${\rm fb}^{-1}$ reach contours from Fig.~\ref{m0mh_3L}.
\label{advantage}}}
\end{center}
\end{figure}

As expected, there is respectable reach beyond LEP-II at small values
of $M_0$ and $\tan\beta$, where the chargino and neutralino 2-body
decays to sleptons are open. As expected from
Fig.~\ref{br_frac_lowtb}(a), the Tevatron has no sensitivity beyond
LEP-II in the region $200\lsim M_0\lsim400$ GeV. For large values of
$M_0$, where the leptonic decays of the gauginos are three-body, we
find some sensitivity for both values of $\tan\beta$. In this region
there is a clear benefit to using optimized cuts. At Run II (2
fb$^{-1}$), with the default set of cuts, only the region with small
$\tan\beta$, small $M_0$, and small $M_{1/2}$ can be explored beyond
LEP-II. With optimal cuts, we see at large $\tan\beta$ a
non-negligible region can be excluded beyond LEP-II. Looking beyond
Run II, we notice that at TeV33 (30 fb$^{-1}$) optimization can prove
equivalent to doubling and sometimes even tripling the total
integrated luminosity!  In Fig.~\ref{advantage} we
show the difference in the required luminosity when using the fixed
set of cuts from \cite{BK} and the optimized cuts.  As a guideline, we
also show the 30 ${\rm fb}^{-1}$ optimal-cut contours (dotted lines) from
Fig.~\ref{m0mh_3L}.

\begin{figure}[t]
\centerline{\psfig{figure=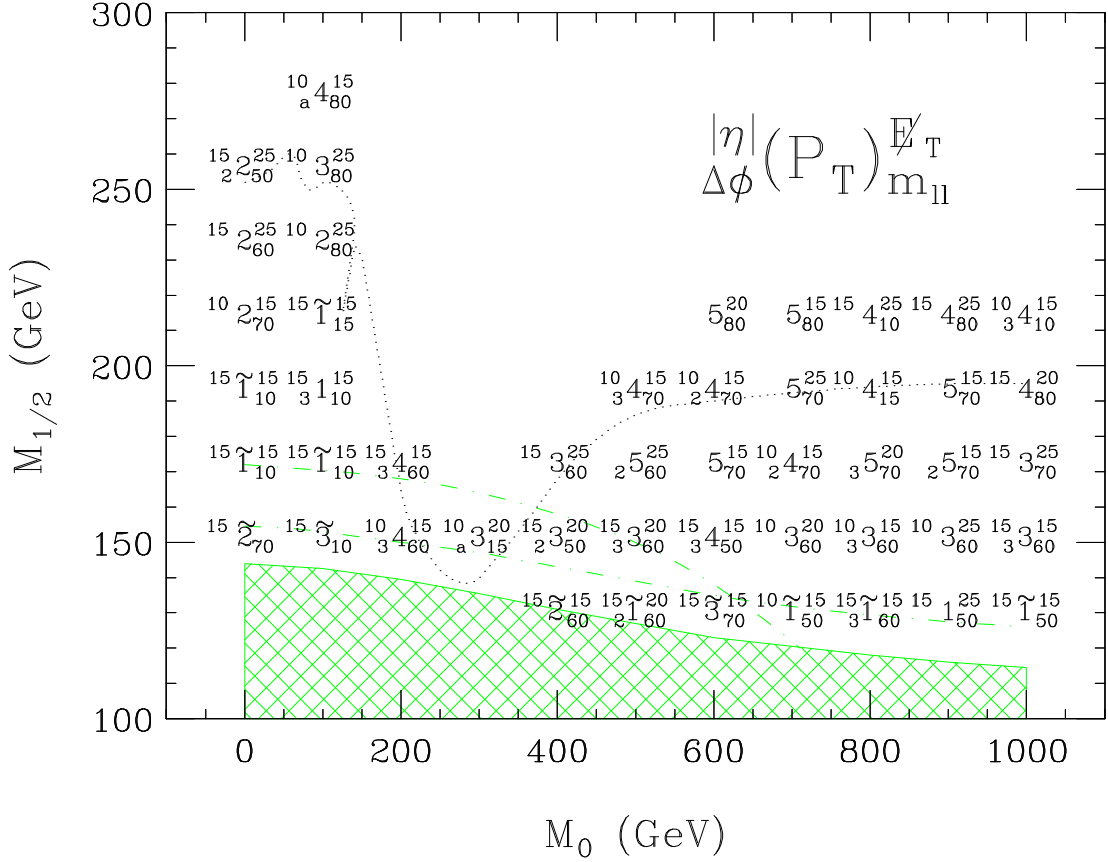,height=4.5in}}
\begin{center}
\parbox{5.5in}{
\caption[] {\small The optimal sets of 3L cuts in the $M_0,\ M_{1/2}$
plane, for $\tan\beta=5$.  The key indicates which symbols correspond
to which cuts (central lepton rapidity, $\met$, invariant mass,
$\Delta\varphi$, and $P_T$ cuts) (see text). The dotted line indicates
the optimal-cut 30 ${\rm fb}^{-1}$ reach contours from
Fig.~\ref{m0mh_3L}.
\label{best cuts 1}}}
\end{center}
\end{figure}

\begin{figure}[t]
\centerline{\psfig{figure=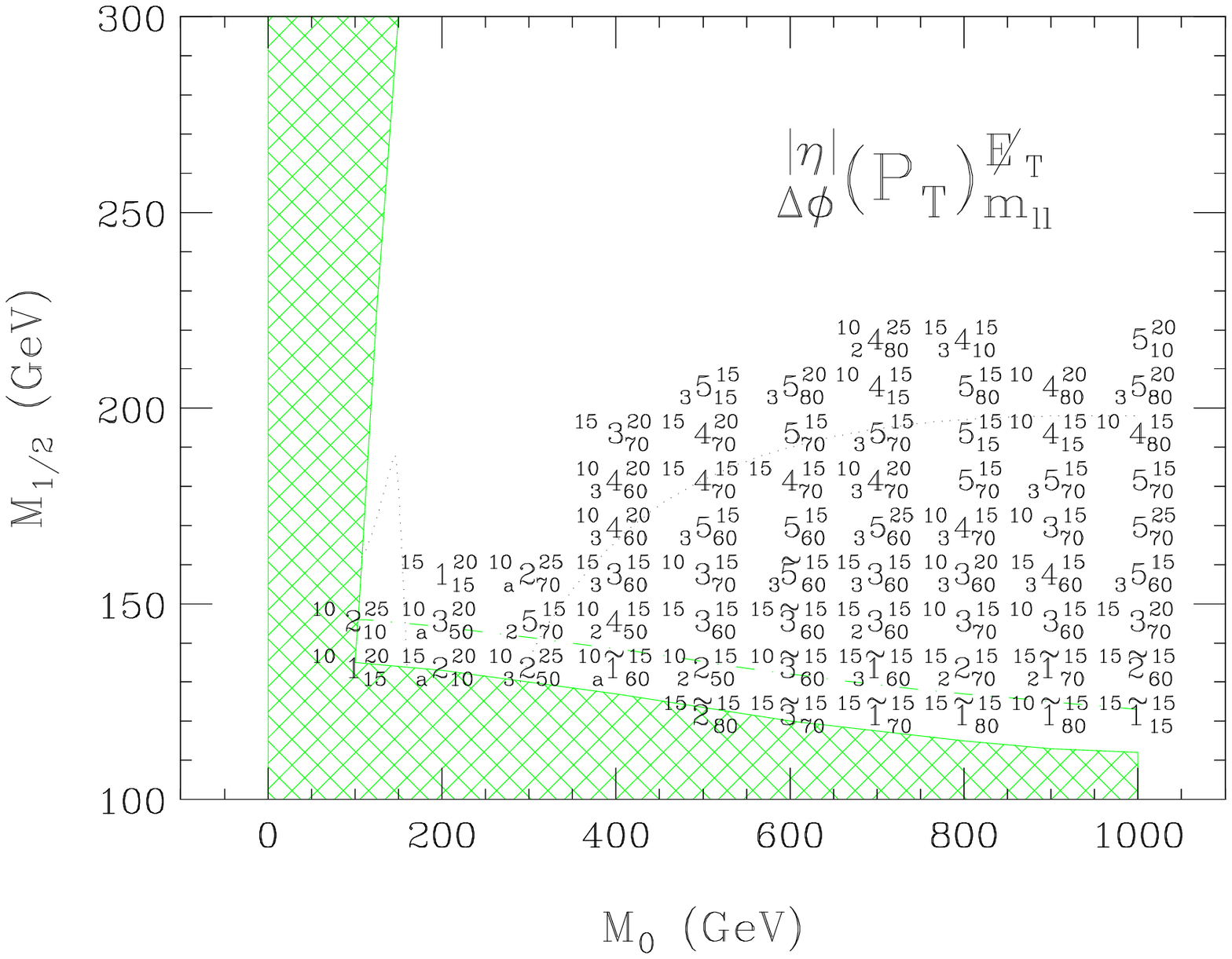,height=4.5in}}
\begin{center}
\parbox{5.5in}{
\caption[] {\small Same as Fig.~\ref{best cuts 1}, with $\tan\beta=35$.
\label{best cuts 2}}}
\end{center}
\end{figure}

It is instructive to examine the optimized sets of cuts.  In
Fig.~\ref{best cuts 1} (Fig.~\ref{best cuts 2}) we show the optimal
sets of cuts for the 3L channel in the $M_0$, $M_{1/2}$ plane, for
$\tan\beta=5$ ($\tan\beta=35$).
We use the following notation to describe the set of cuts at each
point. The central symbol is the number of the $p_T$ cut according to
Table~\ref{pt cuts}.  The left superscript indicates which central
lepton $\eta$ cut was chosen, either $|\eta|<1.0$ (labeled ``10'') or
$|\eta|<1.5$ (``15''). A left subscript denotes the type of
$|\Delta\varphi|$ cut: for the two highest $p_T$ opposite sign, same
flavor leptons the $|\Delta\varphi|<2.5$ cut is indicated by ``2'',
the $|\Delta\varphi|<3.0$ cut by ``3'', and no symbol indicates no
$|\Delta\varphi|$ cut. The cut $|\Delta\varphi|<2.5$ for {\em any}
pair of opposite sign, same flavor leptons is indicated by ``a''.  The
right superscript shows the $\met$ cut: $\met>\{15,20,25\}$ GeV
(``15'',``20'',``25''), or no cut (no symbol).  A right subscript
denotes the dilepton invariant mass cut:
$|m_{\ell^+\ell^-}-M_Z|>\{10,15\}$ GeV (``10'',``15'') or
$|m_{\ell^+\ell^-}|<\{50,60,70,80\}$ GeV
(``50'',``60'',``70'',``80''). And finally, a tilde over the central
symbol indicates that the luminosity limit came from requiring 5
signal events rather than $3\sigma$ exclusion.

The jet veto cut is not indicated on the figures. However, except for
one point at large $\tan\beta$ the jet veto was never selected as an
optimal cut. Indeed, the major 3L background events (from WZ) are just
as likely to contain extra jets as the signal.

There are several lessons to be drawn from Figs.~\ref{best cuts 1} and
\ref{best cuts 2}. As expected, in those cases where the signal is
strong and dominates over the background (typically for small
$M_{1/2}$ or small $M_0$), softer cuts are beneficial.  Indeed, the
majority of the points which can be discovered with 5 signal events,
have selected $p_T$ cut ``1'', which is the most lenient set of $p_T$
cuts. For larger values of $M_{1/2}$, where the background is more
important, harder cuts on the lepton $p_T$'s are preferred. In fact,
we see that in the region of interest for TeV 33 at large $M_0$, the
hardest $p_T$ cuts we consider (from Ref.~\cite{BPT}, indicated by
``5''), often work best.

Figs.~\ref{best cuts 1} and \ref{best cuts 2} clearly indicate the
advantage of the invariant mass cuts introduced in
Section~\ref{sec:analysis}. Of all the points on the plots where the
background is an issue, there are only a few at which a conventional
$Z$-mass window cut is optimal. In all other cases it is advantageous
to cut all events with invariant dilepton masses above a certain
threshold.  Notice also how the value of the threshold tends to
increase with increasing $M_{1/2}$. This is expected because the sharp
edge in the signal distribution is roughly at $0.4\cdot M_{1/2}$
(see Fig.~\ref{inv mass}).

We also see from the figures that a missing $E_T$ cut is preferred
essentially everywhere in parameter space. However, a soft $\met$ cut
($\met>15$ GeV) often gives the better reach. The latest trilepton
analyses \cite{BPT,BK} have chosen to {\em increase} the $\met$ cut
from its nominal Run I value of 20 GeV to 25 GeV.  Our results suggest
that in the off-line analysis one is better off with a lower $\met$
cut. However, more work is needed to conclusively determine what the
best $\met$ cut will be in the actual analysis. For example, triggering
and energy mismeasurement issues will have to be carefully taken into
account.

Lastly, in the great majority of parameter space it is better not to
require a $\varphi$ cut.

Looking back at Fig.~\ref{m0mh_3L}(a), we observe that near the
slepton-chargino mass thresholds the reach becomes diluted. This
effect can easily be overlooked, since it only shows up very close to
threshold.  We find that it is entirely due to the suppressed signal
acceptance.  Indeed, although the branching ratio is increased
immediately below threshold, the lepton resulting from the
$\tilde\chi_2^0\rightarrow \tilde \ell^\pm \ell^\mp$ decay tends to be
very soft and it can fail the analysis cuts.

\begin{figure}[t]
\centerline{\psfig{figure=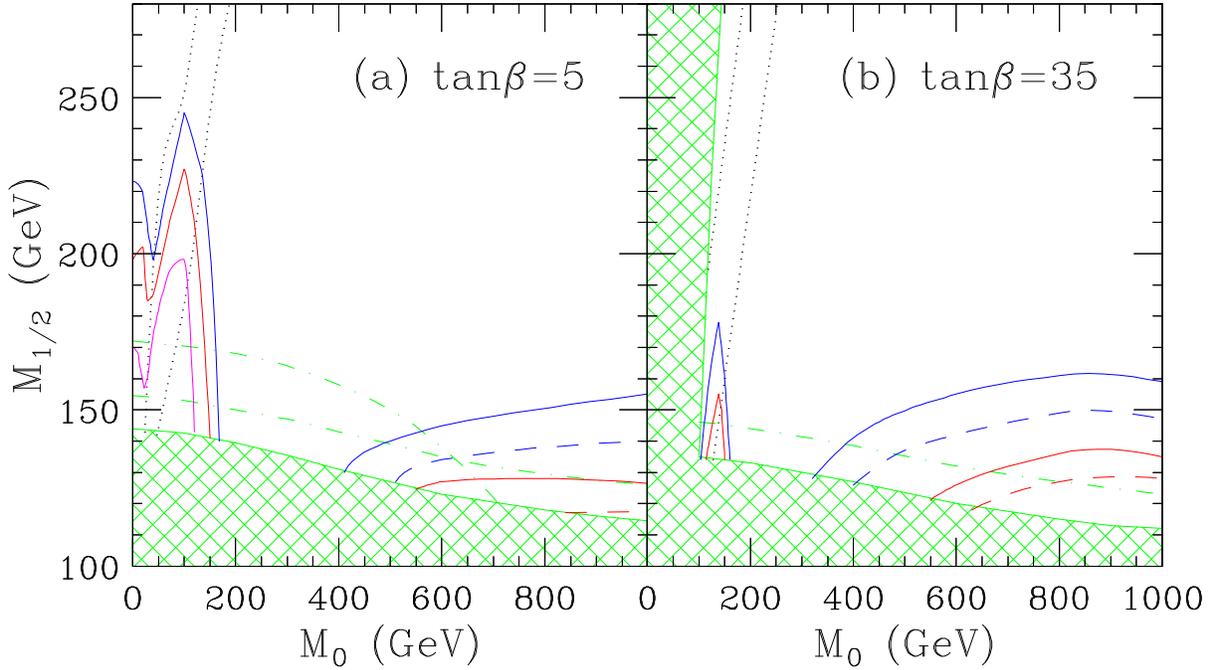,height=3.5in}}
\begin{center}
\parbox{5.5in}{
\caption[] {\small The same as Fig.~\ref{m0mh_3L}, for the 2L channel.
The dashed lines correspond to a set of cuts from Ref.~\cite{JN}.
The 2, 10 and 30 fb$^{-1}$ reach contours are plotted.
\label{m0mh_2L}}}
\end{center}
\end{figure}

We next discuss the prospects for the 2L channel. In
Fig.~\ref{m0mh_2L} we show the 2L channel reach for the Tevatron.
Due to the much larger background the reach in the 2L channel is not
as great as in the 3L channel.
There is one important exception, however -- the 2L channel does not
lose sensitivity near the chargino-charged slepton threshold.  Indeed,
because of the Majorana nature of the neutralinos, in 50\% of the
events the lost soft lepton in the $\tilde\chi_2^0\rightarrow \tilde
\ell^\pm \ell^\mp$ decay is of the {\em opposite} sign as the charged
lepton from the chargino decay. The remaining two hard leptons are
then of the same sign and they are readily reconstructed. Therefore,
near the $\tilde\ell/\tilde\chi_1^\pm$ mass threshold, the 2L
channel may prove to be a valuable alternative to 3L.

Fig.~\ref{m0mh_3L}(b) reveals that at large $\tan\beta$ one starts to
lose 3L sensitivity in the small $M_0$ region, since the decays to tau
final states dominate.  In fact, for $M_0\lsim 300$ GeV and
$\tan\beta=35$, we find no 3L reach in Run II beyond LEP-II. Only with
multiple years of running and collecting soft lepton events will the
Tevatron be able to start improving on the LEP mSUGRA bounds. One can
improve this situation by considering alternative signatures with tau
jets \cite{LM}. Of those, the 2L1T channel is singled out on the basis
of both sensitivity and statistical importance.

\begin{figure}[t]
\centerline{\psfig{figure=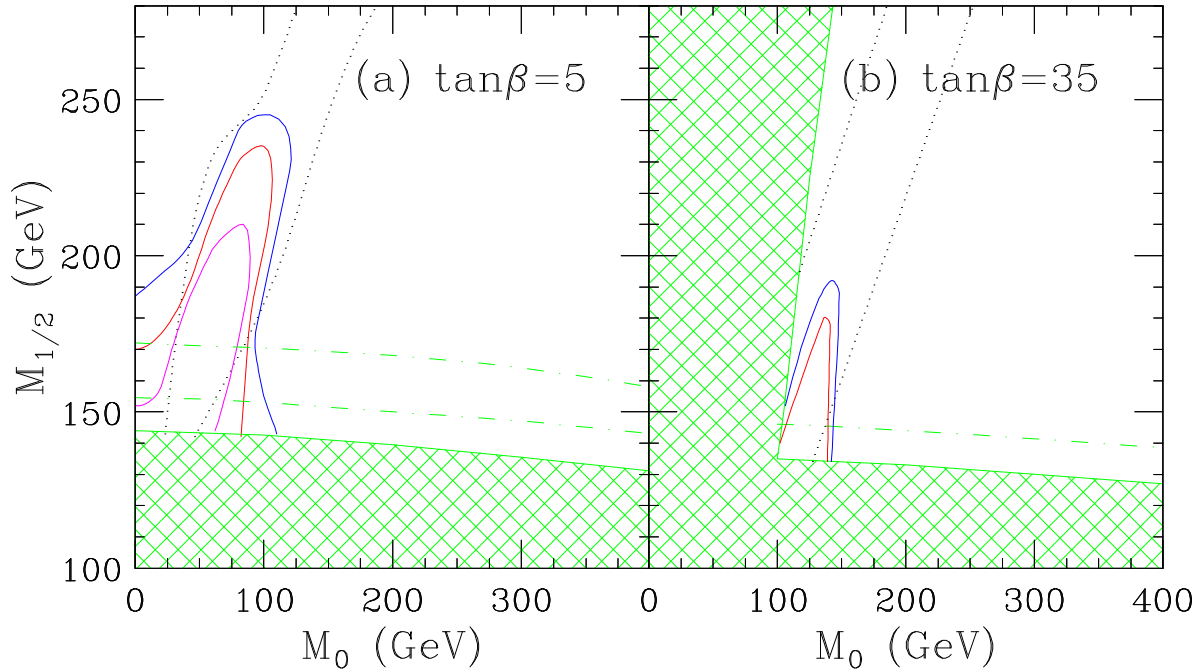,height=3.5in}}
\begin{center}
\parbox{5.5in}{
\caption[] {\small The same as Fig.~\ref{m0mh_3L}, for the 2L1T
channel. In this case the optimal cuts yield very little improvement
relative to a default set of cuts from Ref.~\cite{LM} (listed in
Sec.~\ref{sec:summary}), so the contours corresponding to the default
cuts are not shown.
\label{m0mh_2L1T}}}
\end{center}
\end{figure}

In Fig.~\ref{m0mh_2L1T} we show the Tevatron reach in the 2L1T channel.
In this case we compare the optimal reach to a set of cuts from
Ref.~\cite{LM} (listed in Sec.~\ref{sec:summary}). We find very little
difference between this fixed set of cuts and the optimal set, so we
do not show the fixed cut lines in the figure.

The 2L1T channel has no reach in the large $M_0$ region. However, when
the two-body decays to staus open up the 2L1T branching fraction is
larger than the other two channels, leading to some sensitivity. While
the region accessible in the 2L1T channel at small $\tan\beta$ is not
competitive with the 3L and 2L reach, it can improve the statistical
significance in case of exclusion, or it can serve as an important
confirmation and provide unique information about the model
parameters in case of discovery.  This channel offers the greatest
reach at large $\tan\beta$ (together with the related much cleaner
signature of two like-sign leptons plus a tau jet \cite{LM}) in the
small $M_0$ region.

\section{Conclusions} \label{sec:conclusions}

In this paper we studied three of the cleanest and most promising
channels for SUSY discovery at the Tevatron in Run II.  We revisited
the trilepton and like-sign dilepton analyses, improving them in
several key aspects. For example, we used a more realistic simulation
of the major backgrounds. We found larger backgrounds than previous
analyses. We used a procedure relying on Run I data to estimate
backgrounds involving fake leptons. And we introduced an invariant
mass cut which took advantage of a sharp edge in the signal dilepton
invariant mass distribution. This cut was generally more effective in
increasing signal-to-noise than the standard invariant mass cuts.
Also, we varied the cuts at each point in the supersymmetric parameter
space, and determined at each point the set of cuts which yields the
largest reach. We found that this cut optimization can significantly
enhance the Tevatron reach.  Lastly, we analyzed the reach of the 2L1T
channel.

If nature is supersymmetric at low energies, and the superpartners (in
particular the gauginos) are light, there is a good chance that the
Tevatron will discover them in its upcoming runs. However, the 3L
signature has limited reach, and we can improve the reach by
optimizing cuts and considering alternative clean signatures.

\section*{Acknowledgments}

We thank J.~Campbell, R.K.~Ellis, J.~Lykken, J.~Nachtman and F.~Paige
for useful discussions.  K.T.M. (D.M.P.) is supported by Department of
Energy contract DE-AC02-76CH03000 (DE-AC02-98CH10886).

{\em Note added:} We thank H.~Baer for providing us with a preliminary
version of a preprint by H.~Baer, M.~Drees, F.~Paige, P.~Quintana and
X.~Tata. This paper is an update of their previous study \cite{BPT} of
the reach of the Tevatron in the 3L channel. Part of the improvement
is the inclusion of the $Z$-width in the $WZ$ background determination.

\end{document}